\documentclass[12pt,preprint]{aastex}

\begin{document}

\title{The Environments of SGRs: A Brief \& Biased Review}

\author{Stephen S. Eikenberry\altaffilmark{1}}

\altaffiltext{1}{Astronomy Department, 212 Space Sciences Building, Cornell University, Ithaca, NY  USA}

\begin{abstract}

	I review some recent developments in our understanding of the
environments of soft gamma-ray repeaters.  I pay particular attention
to the apparent association betwwen SGR 1900+14 and SGR 1806-20 and
embedded clusters of stars.

\end{abstract}

\section{Introduction}

	The environments of soft gamma-ray repeaters (SGRs) can
provide many important insights into these intriguing and unusual
objects.  First, the environment of an SGR may provide clues to its
origins, including possible progenitor stars, the age of the SGR, its
space velocity, etc.  Such environmental clues may be gleaned from a
surrounding supernova remnant, nearby stars, and unusual ISM features,
among other things.  Secondly, by studying the impact of the SGR on
its surrounding ISM (or vice versa) we can also hope to learn more
about the evolution of SGRs.

	In this review, I will first go over what we knew (or thought
we knew) about SGR environments previously.  I will then go over our
current state of knowledge regarding the environments of SGRs which
are well-localized at the time of the Woods Hole 2001 meeting: SGR
0525-66, SGR 1900+14, and SGR 1806-20.

\section{What We Used to ``Know''}

	Prior to about 1998, the community had a reasonably clear
(though not universally accepted) picture of SGR environments.  The
three well-localized SGRs were all ``known'' to be associated with
supernova remnants.  SGR 0525-66 lies inside the supernova remnant N49
in the LMC \citep{Rothschild}.  SGR 1900+14 was localized to within a
few arcminutes of G42.8+0.6 (\citet{Hurley94}; \citet{Vasisht94}), and
also apparently coincident with an unusual double MIa star system
\citep{Vrba96}.  Finally, SGR 1806-20 was found to lie very close to
the apparent SNR G10.0-0.3 \citep{Hurley94}.  This supposed SNR had a
bright, variable radio core apparently associated with an X-ray source
(\citet{Vasisht}; \citet{KF93}), indicating a possible plerionic SNR
similar to the Crab Nebula.  Subsequent observations revealed a bright
heavily-absorbed infrared star at the radio core position
\citep{Shri95}, which was then found to be a candidate luminous blue
variable (LBV) star based on its IR spectrum \citep{vK95}.  Due to the
extreme rarity of both LBVs ans SGRs, a chance superposition of these
two seemed very unlikely and led to the conclusion that the SGR must
be related to the LBV.  As I describe below, we now recognize that
many of these associations are either coincidental or due to some more
complicated underlying associations.

\section{SGR 0525-66}

	Since the giant outburst of March 1979, SGR 0525-66 has been
associated with the Large Magellanic Cloud \citep{Mazets}.  More
recently, \citet{Rothschild} found an X-ray source within the
$\gamma$-ray error box, which lies inside the contours of the N49 SNR.
Simple analyses show that the chance projection of the SGR within
this SNR is $<1 \%$, even making rather stringent assumptions
\citep{Gaensler}.  Thus, this SGR/SNR association seems robust.

	SGR 0525-66 and its host SNR also have the lowest extinction
of any known SGR, with $A_V \sim 1$ mag.  More recently,
\citet{Kaplan01a} have obtained HST optical observations of the area
surrounding SGR 0525-66.  They find several stars within the X-ray
positional error box, as well as some apparent bright ``lumps'' from
the SNR itself.  The stars are consistent with G/K main sequence
stars.  It is interesting to note, in comparison with the other
well-localized SGRs, that there are {\bf no luminous stars nearby}.

\section{SGR 1900+14}

	The $\gamma$-ray IPN localization for SGR 1900+14 initially
led to an apparent association with an X-ray source \citep{Hurley94}.
This in turn led to an initial optical/IR identification of SGR
1900+14 with an unusual pair of M-type supergiant stars
\citep{Vrba96}.  The relative scarcity of such pairings made a
coincidental alignment of the SGR with them moderately unlikely.

	However, the 1998 major outburst of SGR 1900+14 produced,
among other things, a radio transient which \citet{Frail99} localized
to $\sim 1$-arcsec precision.  This position was shown to be
inconsistent with the M-star pair \citep{Eiken01} (see Figure 1), and
no variable source was found from IR observations during the 1998
flare (\citet{Opp}; \citet{Eiken01}).

	More recently, \citet{Vrba00} have shown that the M-star pair
are in fact the two brightest members of an entire cluster of
supergiant and giant stars.  This dense cluster has a reddening
consistent with the X-ray absorption towards SGR 1900+14, and, at an
assumed distance of $\sim 10-15$ kpc, lies a projected distance of
$\sim 1$ pc from the SGR.  The same arguments apply for the
association with the SGR and the cluster -- the odds of having two
such unusual objects coincidentally aligned on the sky are low.  One
possible interpretation is that SGR 1900+14 was ``born'' in the
cluster of stars, and either left in its progenitor stage or was
ejected by a supernova kick (or both).  If we assume this scenario and
a ``standard'' pulsar kick velocity of a few hundred $\rm km/s$, then a
separation of 1 pc implies an age for SGR 1900+14 of $<10^4$ years --
consistent with several models for SGR activity.

\section{SGR 1806-20}

	As noted above, SGR 1806-20 was thought to be associated with
a luminous blue variable (LBV) star which lies at the time-variable
(in both flux and morphology) core of the radio nebula G10.0-0.3
(\citet{Vasisht}; \citet{KF93}).  However, the recent Inter-Planetary
Network (IPN) localization of SGR 1806-20 provides a position
inconsistent with that of the LBV star and radio core
\citep{Hurley99c}.  Furthermore, \citet{Gaensler} argue that G10.0-0.3
is not a supernova remnant at all, but is rather powered by the
tremendous wind of the LBV star.  Infrared observations of the field
of SGR 1806-20 reveal that the LBV star is not alone, but appears to
be part of a cluster of embedded, hot, luminous stars \citep{Fuchs},
and the IPN position for SGR 1806-20 is consistent with membership in
that cluster.  Recently,
\citet{LBV} have used near-infrared photometry and spectroscopy to
conclude that this cluster contains what may be the most luminous star
in the Galaxy (the LBV star), at least one Wolf-Rayet star of type
WCL, and at least two blue ``hypergiants'' of luminosity class Ia+.
These properties make the cluster resemble a somewhat smaller and
older version of the ``super'' star cluster R136 \citep{Massey},
making the potential association with SGR 1806-20 even more
intriguing.

	{\it Chandra} observations of SGR 1806-20 have provided a
sub-arcsecond localization of the SGR in this crowded field
(\citet{Kaplan}; \citet{Eiken02}).  The localization matches the IPN
position for SGR 1806-20, and completely excludes a direct association
of the SGR with the LBV and radio core.  This seems to confirm the
conclusion of \citet{Gaensler} that G10.0-0.3 is in fact not a SNR,
but is instead a radio nebula powered by the mass-loss wind of the LBV
star at its core.  Figure 2 shows J-, H-, and K-band infrared images
taken with OSIRIS on the CTIO 4-m telescope of the region near SGR
1806-20, along with the 90\% positional error circle.  As noted above,
SGR 1806-20 lies in the direction of an unusual embedded cluster of
massive, luminous young stars (\citet{Fuchs};
\citet{LBV}), with a distance of $14.5 \pm 1.4$ kpc and a reddening of
$A_V = 29 \pm 2$ mag (\citet{Corbel};
\citet{LBV}).  The fact that stars D and E in Figure 2 have $J-K =
5.0$ mag indicates that they are members of this cluster ($E_{J-K} =
5.0$ mag for $A_V = 29$ mag), and thus that SGR 1806-20 lies within
the radial extent of the cluster on the sky.  Furthermore, the X-ray
absorption towards SGR 1806-20 is $\sim 5-6 \times 10^{22} \ {\rm
cm^{-2}}$ (see above, and
\citet{Mereg}), which is consistent with the extinction towards the
cluster.  Thus, it seems likely that SGR 1806-20 is also a member of
this massive star cluster at a distance of $14.5 \pm 1.4$ kpc.

	The IR colors (and upper limits) of the two candidate
counterparts to SGR 1806-20 are consistent with both of them being
stellar members of the star cluster ($J-K = 5.0$ mag, $H-K = 2.0$
mag).  Note that the brighter star just outside the error circle
(``C'' in Figure 2) appears to be a foreground star ($J-K = 3$ mag),
confirming that it is not a likely counterpart to the SGR.  However,
as can been seen in Figure 2, the field of SGR 1806-20 is highly
crowded in the IR by both foreground/background objects and cluster
members, and the simple fact that two stars lie within the 90\%
confidence error circle and another just outside it shows that the
probability of a chance coincidence of unrelated IR objects is high.
Thus, we cannot conclude definitely that {\it either} of the possible
counterparts is actually related to SGR 1806-20.  If both candidates
are in fact members of the cluster, we can estimate their absolute
magnitudes to be $M_K = -2.3$ mag (A) and $M_K = -0.4$ mag (B).  For
Star A, this is consistent with stars of luminosity matching a B1V or
K3III star, and is inconsistent with any stars of luminosity class I.
After correcting for the $H-K = 2$ mag differential extinction toward
the cluster, the intrinsic color of $(H-K)_{intrins} = 0.8 \pm 1.2$
mag is essentially consistent with all stellar spectra earlier than
late M, and does not significantly constrain the classification.  For
Star B, the absolute magnitude is consistent with stars of luminosity
matching a B8V star, and is inconsistent with any stars of luminosity
class III or higher.  It is important to note that no observations
have yet probed the distribution of stars in the cluster with masses
below that of late B main sequence stars.  Thus, it is possible that
there are even further stars within the error circle at significantly
lower mass/luminosity, unless the cluster mass distribution shows a
sharp lower cutoff.

	One particularly intriguing aspect of the association between
SGR 1806-20 and the star cluster is that a neutron star progenitor
went supernova {\it before} the stars currently observed in the
cluster.  Since more massive stars evolve to the supernova stage more
rapidly, {\it if} the progenitor of SGR 1806-20 formed at the same
time as the currently observed massive stars, its mass must have been
greater.  However, the mass estimate for the LBV is $>200 M_{\odot}$
\citep{LBV}, and several of the other stars are likely to have masses
in the range of $\sim 50-100 M_{\odot}$ \citep{LBV}.  While recent
theories have predicted that very massive stars may produce neutron
star remnants due to the effect of envelope loss via dense stellar
winds, the upper limit on their masses is $\sim 80 M_{\odot}$, with
higher mass stars producing massive black hole remnants.  Thus, it
seems unlikely that the SGR 1806-20 progenitor formed at the same time
as the currently observed massive stars in the cluster.

	Alternately, SGR 1806-20 may have formed prior to these stars
-- in fact, \citet{Kaplan} suggest that the supernova event that
produced SGR 1806-20 may have triggered the star formation activity
that produced the massive stars in the cluster.  However, the massive
stars we observe in the cluster are evolved, with ages of $\sim 10^6$
yr to reach the LBV and Wolf-Rayet stages of their lives.  If the SGR
1806-20 supernova event led to the birth of these stars, then SGR
1806-20 is much older than the $\sim 10^3 -- 10^4$ yr typically
considered for magnetars.  Alternately, SGR 1806-20 may simply be
taken as evidence for prior massive star formation at this location.
While at least one supernova occurred here, perhaps it was not the
first, and an earlier supernova event triggered the formation of the
currently observed massive stars.

\section{Some Conclusions}

	Here I present some of the conclusions of the above discussions:

\begin{itemize}

\item Of the well-localized SGRs, only SGR 0525-66 seems to have a clear association with a SNR.

\item SGR 1900+14 is not {\it directly} associated with the MIa double star system.

\item SGR 1806-20 is not {\it directly} associated with the LBV star near it, nor is there a SNR evident in this neighborhood

\item Both SGR 1806-20 and SGR 1900+14 are in/near dense clusters of massive stars.  These are rare enough that a chance superposition is low.

\item The association of SGRs with such clusters explain the apparent, but eventually false, association with some of the particular massive stars in the clusters.

\item SGR 0525-66 is {\it not} associated with any apparent cluster of massive stars.

\item Th environments of SGRs are giving some very interesting, if somewhat mixed, signals regarding the origins and evolutions of these objects.

\end{itemize}

\begin{figure}
\plotone{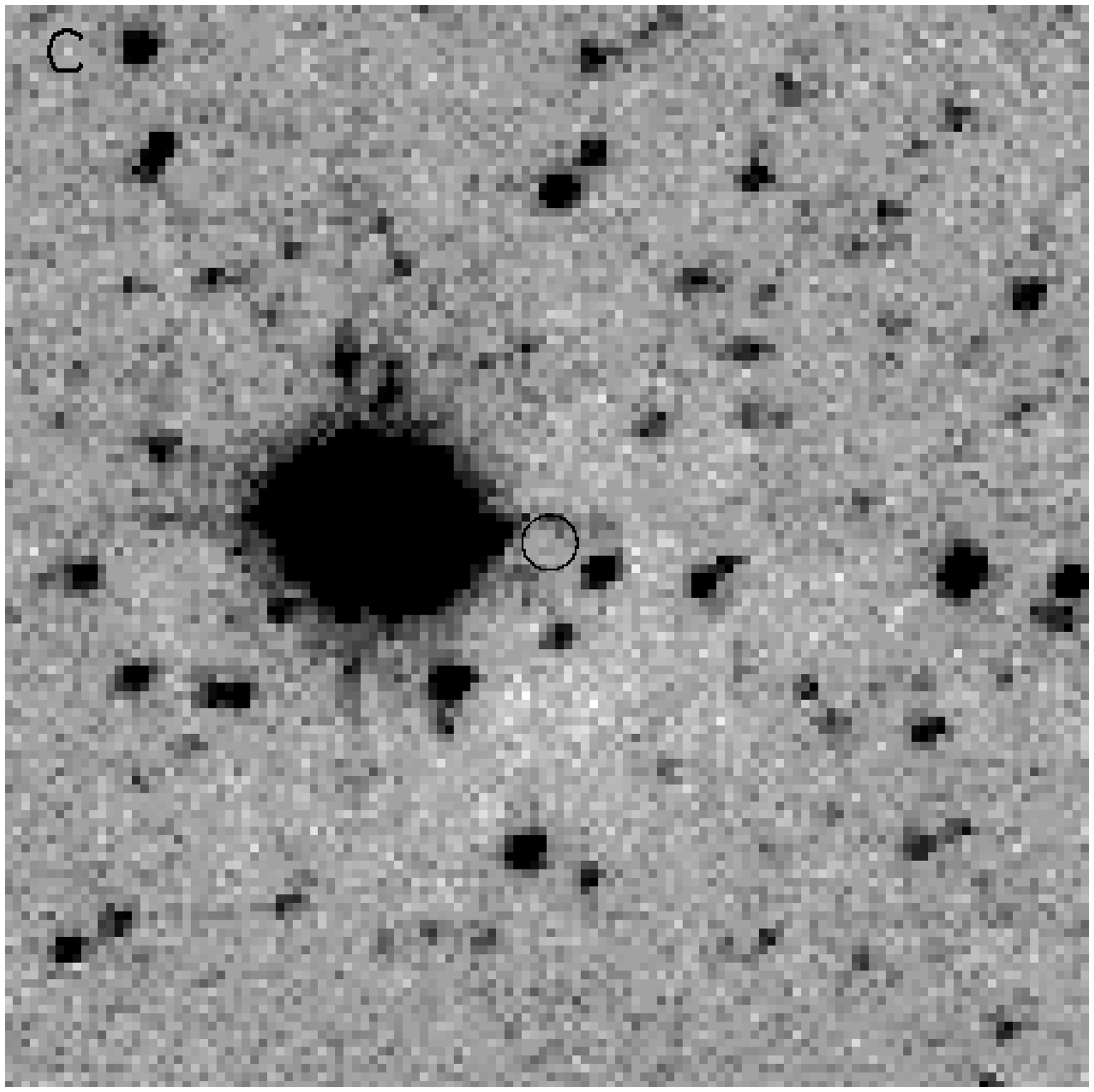}
  \caption{Infrared K-band image of the field near SGR 1900+14 taken
  near the outburst of 1998 (after \citet{Eiken01}).  The small circle
  indicates the radio positional error box for SGR 1900+14.  The
  cluster of stars including the luminous M-star pair are seen off to
  the left.  North is up and east is to the left in this image.}
\end{figure}

\begin{figure}
{\plottwo{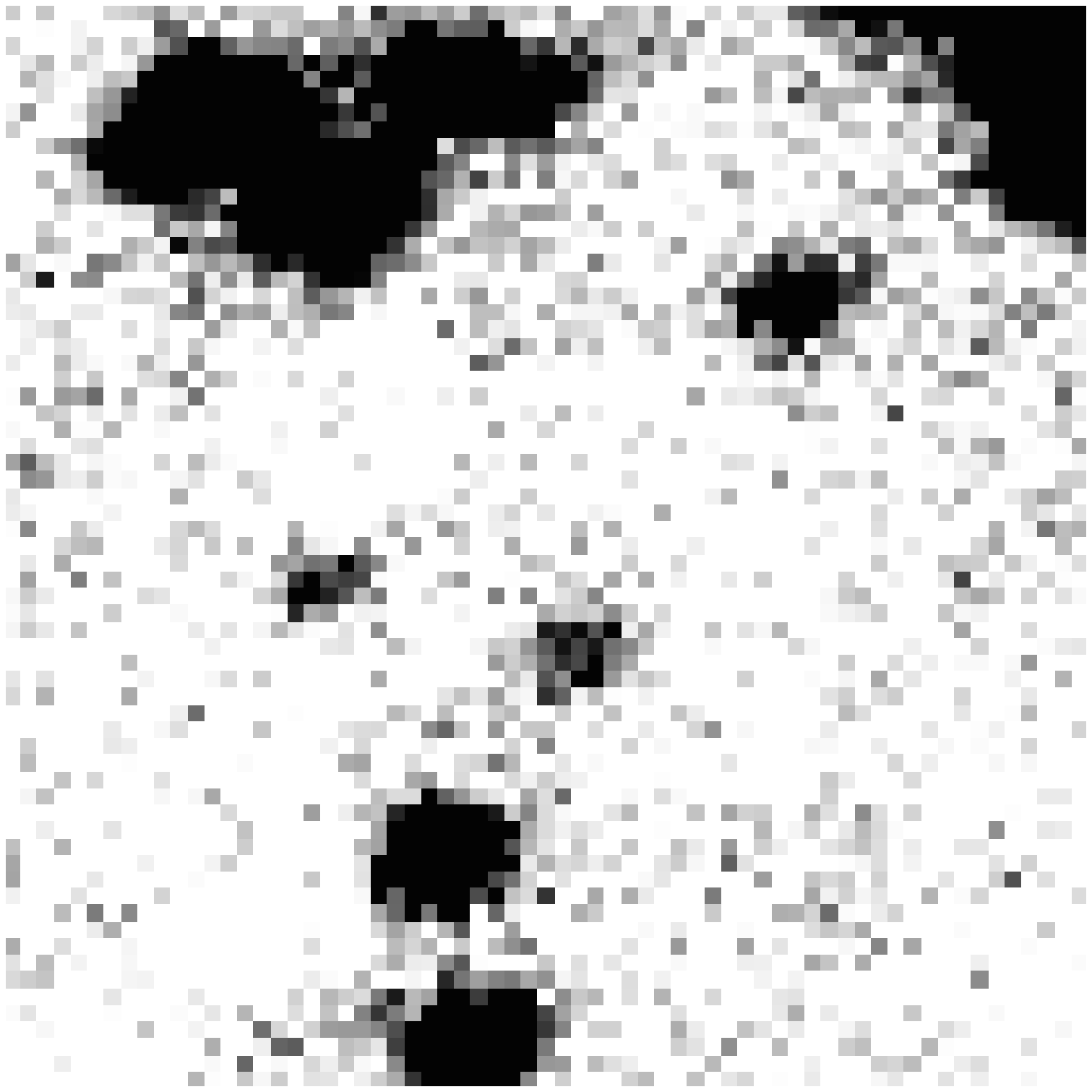}{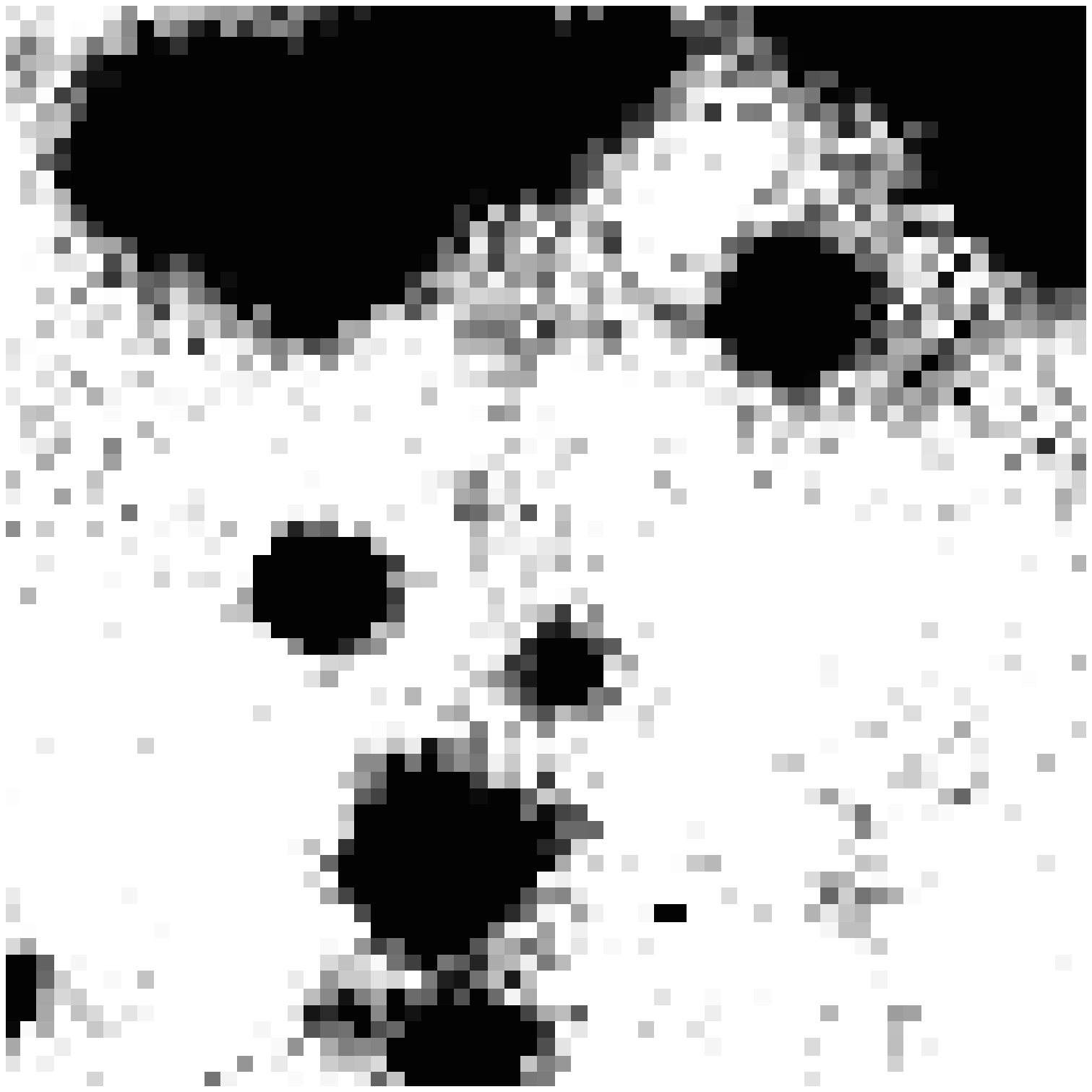}}
\vskip 2.0 mm
{\plottwo{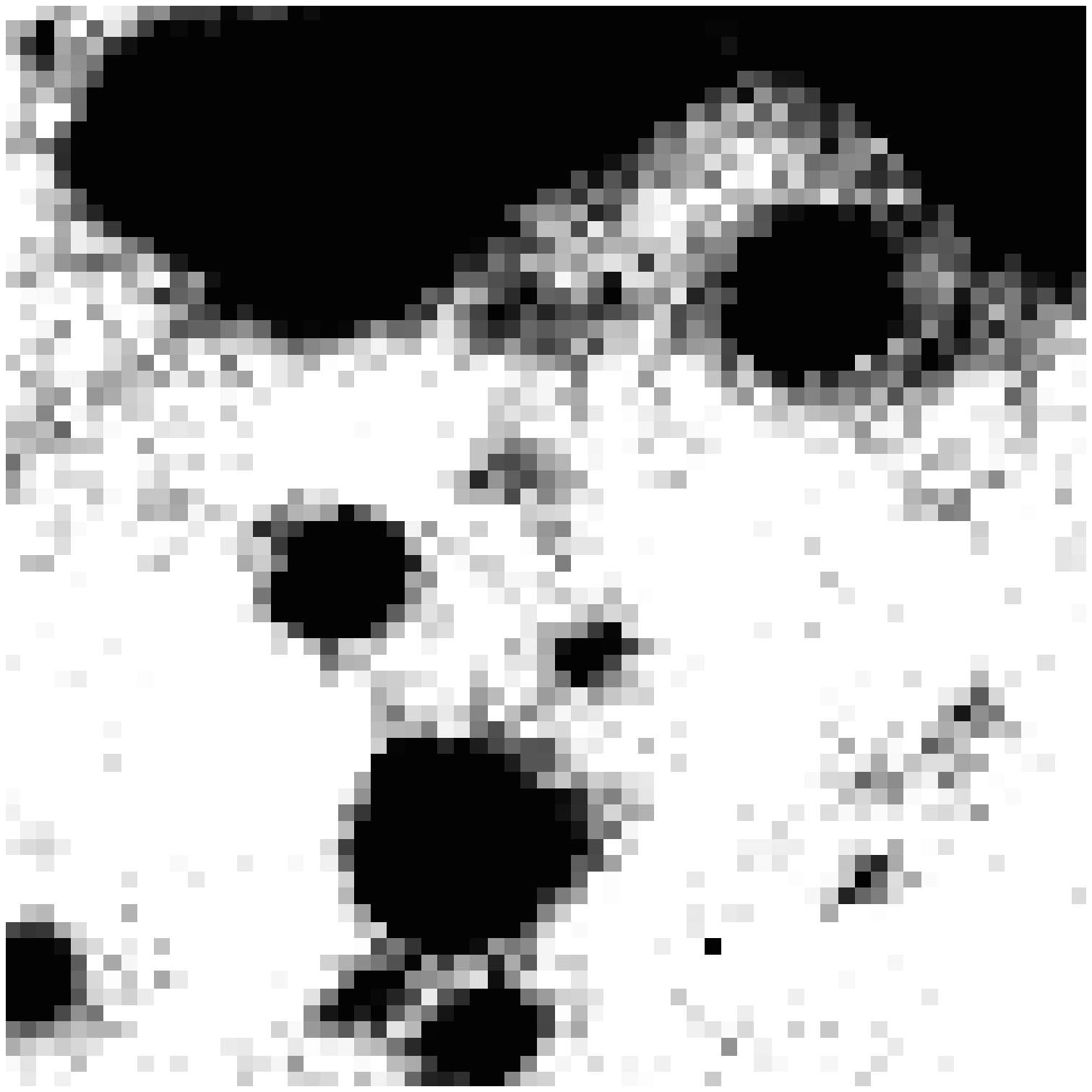}{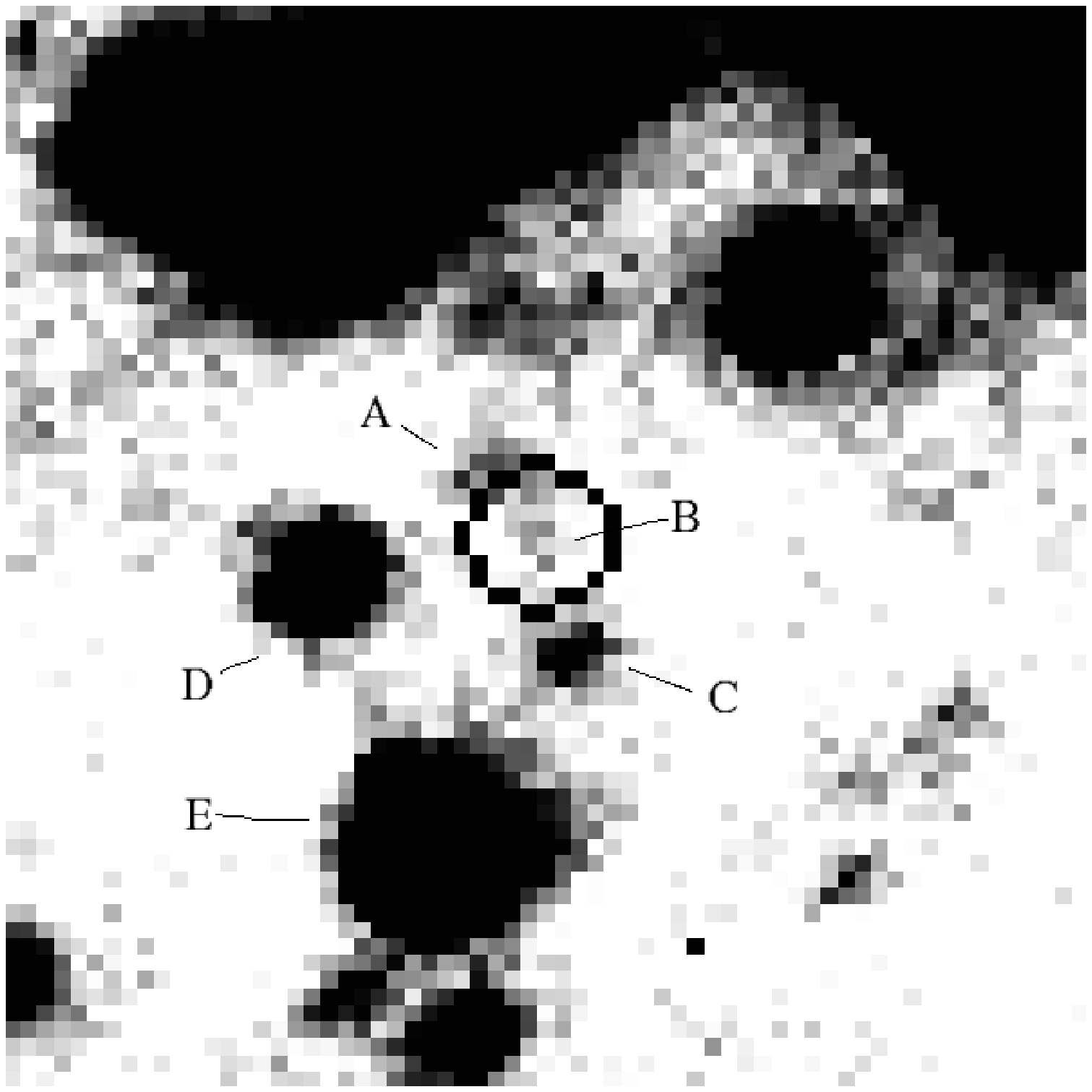}}

  \caption{Infrared images of the field of SGR 1806-20 (as in \citet{Eiken02}).  Images shown are: J-band (top left); H-band (top right); K-band (bottom left); K-band with stars labelled (bottom right).  North is up and east is to the left.  The circle indicates the 0.7-arcsec-diameter X-ray error circle from {\it Chandra}.  The bright stars at the top of the images are near the cluster center.}

\end{figure}

\end{document}